\documentclass[aps,prd,preprint,superscriptaddress,nofootinbib]{revtex4-1}
\usepackage{graphicx,xcolor,amsmath}
\usepackage{hyperref}
\usepackage{bm}
\usepackage{mathtools}
\usepackage{siunitx}
\usepackage{booktabs}
\usepackage{multirow}
\usepackage{colortbl}
\usepackage[capitalise]{cleveref}

\usepackage[parfill]{parskip}

\newcommand{\mbctwos}{\ensuremath{m_{B_c^+}(2S)} }
\newcommand{\mbctwosp}{\ensuremath{m_{B_c^{*+}}(2S)} }

\newcommand{\mbcones}{\ensuremath{m_{B_c^{*+}}(1S)} }

\newcommand{\bctwos}{\ensuremath{B_c^+(2S)} }
\newcommand{\bctwosp}{\ensuremath{B_c^{*+}(2S)} }
\newcommand{\bcones}{\ensuremath{B_c(1S)} }
\newcommand{\bconesp}{\ensuremath{B_c^{*+}(1S)} }

\newcommand{\fbcs}{\ensuremath{f_{B_c^{*+}}(2S)} }
\newcommand{\fbc}{\ensuremath{f_{B_c^+}(2S)} }

\newcommand{\fbconesp}{\ensuremath{f_{B_c^{*+}}(1S)} }
\newcommand{\fbcones}{\ensuremath{f_{B_c^+}(1S)} }

\newcommand{\fbctwosp}{\ensuremath{f_{B_c^{*+}}(2S)} }
\newcommand{\fbctwos}{\ensuremath{f_{B_c^+}(2S)} }

\begin{document}

\title{Properties of excited $B_c$ states in QCD}

\author{T.~M.~Aliev}
\email{taliev@metu.edu.tr}
\affiliation{Department of Physics, Middle East Technical University, Ankara, 06800, Turkey}

\author{T.~Barakat}
\email{tbarakat@ksu.edu.sa}
\affiliation{Department of Physics, King Saud University, Riyadh, 11451, Saudi Arabia}

\author{S.~Bilmis}
\email{sbilmis@metu.edu.tr}
\affiliation{Department of Physics, Middle East Technical University, Ankara, 06800, Turkey}

\date{\today}

\begin{abstract}
  The mass and leptonic decay constants of recently observed two new excited $B_c$ states at LHC are studied within the QCD sum rules. Considering the contributions of the ground and radially excited states, the mass and residues of the excited states of pseudoscalar and vector mesons are calculated in the framework of two different approaches of the QCD sum rules, namely, linear combinations of the corresponding sum rules and its derivatives as well as QCD sum rules with the incorporation of the least square fitting method.  The obtained results on mass $\mbctwos = 6.88 \pm 0.03~\rm{GeV} $ and $\mbctwosp = 6.94 \pm 0.03~\rm{GeV} $ are in good agreement with the experimental data. Our predictions for the decay constants of these states are: $\fbc = 0.42 \pm 0.02~\rm{GeV} $ and $\fbcs = 0.46 \pm 0.01~\rm{GeV}$, which can be checked at future experiments to be conducted at the LHC. Comparison of our results with the predictions of the other approaches on mass and residues is also presented.
\end{abstract}

\maketitle

\section{Introduction}
\label{sec:1}
$B_c$-meson system formed from the two heavy quarks with different flavors can be very useful in obtaining information about heavy-quark dynamics and understanding the dynamics of the strong interaction (see~\cite{Eichten:2019gig} and \cite{Gershtein:1994jw} and references therein). The ground state $B_c$-mesons were discovered at CDF and D$\O$ experiments for the first time through the decay channels $B_c \rightarrow J/\Psi l \nu$ and $B_c \rightarrow J/\Psi l \pi$ decay modes~\cite{Aaltonen:2007gv,Abazov:2008rba,Abazov:2008kv,Abe:1998wi}. Later, the discovery of the B-meson was confirmed by the LHCb, CMS and ATLAS Collaborations via various decay modes: $B_c \rightarrow J/\Psi l \nu$~\cite{Aaij:2017tyk} and $B_c \rightarrow J/\Psi l \pi^+$~\cite{Aaij:2013wda}, $B_c \rightarrow J/\Psi D_s^{*}$~\cite{Aaij:2013gia}, $B_c \rightarrow J/\Psi \pi$~\cite{Aaij:2012dd} etc..
Recently, two excited $2^{1}S_0$ and $2^{3}S_1$  states of $B_c$ meson have been observed in the invariant mass spectrum of the $B_c^{+} \pi^+ \pi^-$ by CMS Collaboration at the centre of mass energy $\sqrt{s} = 13~\rm{TeV}$~\cite{Sirunyan:2019osb}. We will denote these states as $B(2S)$ and $B^*(2S)$ from now on. The state \bctwos has mass $m_{B_c^+}(2S) = 6871 \pm 1.2~(\rm{stat}) \pm 0.8~(\rm{sys})~\rm{MeV},$ and the mass difference of these states is $\mbctwosp - \mbctwos = 29 \pm 1.5~(\rm{stat}) \pm 0.7~(\rm{sys})~\rm{MeV}.$ Very recently, LHCb Collaboration announced the observation of these states and obtained that \bctwos state has a mass $m = 6841.2 \pm 0.6~\rm{(stat)} \pm 0.1(\rm{sys})~\rm{MeV}$, and mass difference of \bctwos and \bctwosp is measured to be $31.1 \pm 1.4~(\rm{stat})~\rm{MeV}$~\cite{Aaij:2019ldo} which are excellent agreement with CMS Collaboration.

These experimental discoveries stimulated the theoretical studies for the investigation of the properties of these mesons. The mass and residues, as well as various decay channels  of the ground state $B_c$-meson within different methods were investigated in many studies, such as quark model~\cite{Soni:2017wvy}, lattice theory~\cite{Mathur:2018epb}, QCD sum rules~\cite{Colangelo:1992cx,Aliev:1992vp,Wang:2012kw,Chabab:1993nz,Bagan:1994dy}, Dyson-Schwinger and the Bethe-Salpeter equations approaches~\cite{Chang:2019eob}. In the present work, we calculate the mass and leptonic decay constants of the excited \bctwos and \bctwos mesons within the QCD sum rules. The rest of the paper is organized as follows. In section~\ref{sec:2}, we derive the mass sum rules, including the contributions of the ground and excited states. In section~\ref{sec:3}, with the help of the two different version of the QCD sum rules, i.e. employing the least square method as well as the construction of linear combinations of sum rules and its derivatives, we perform numerical analysis of the sum rules obtained in the previous section and extract the mass and leptonic decay constants of \bctwos and \bctwosp excited states. 
\vfill
\section{Sum rules for the spectroscopic parameters of the excited $B_c$  meson}
\label{sec:2}
To determine the mass and leptonic decay constants of the excited states of $B_c$-mesons within the QCD sum rules approach, we consider following correlation functions at the quark level,
\begin{equation}
  \label{eq:1}
  \begin{split}
      \Pi_1 &= i \int d^4x e^{i q x} \langle 0 | J(x) J(0) | 0 \rangle \\
      \Pi_{\mu \nu} &= i \int d^4x e^{i q x} \langle 0 | J_\mu(x) J_\nu(0) | 0 \rangle \\
                  &= (g_{\mu \nu} - \frac{q_\mu q_\nu}{q^2}) \Pi_2(q^2) + q_\mu q_\nu \Pi_3(q^2)
  \end{split}
\end{equation}
where the $J(x) = (m_b + m_c) \bar{c} i \gamma_5 b$ and $J_\mu(x) = \bar{c} \gamma_\mu b$ are the interpolating currents carrying the same quantum numbers of pseudoscalar and vector mesons, respectively. To obtain the desired sum rules for the mass and residues, the correlation function needs to be calculated in two different domains in $q^2$; in terms of hadrons and with the help of the operator product expansion (OPE) in deep Euclidean region. By matching the results of these representations, one can obtain the corresponding sum rules.
The hadronic representation of the correlation function satisfies the dispersion relation
\begin{equation}
  \label{eq:14}
  \Pi_\alpha(q^2) = \frac{1}{\pi} \int \frac{\rho_\alpha(s) ds}{s-q^2} + \text{ subtractions,}
\end{equation}
where $\alpha = 1(2)$ corresponds to $\Pi_1 (\Pi_2)$, and $\rho_\alpha(s)$ is the corresponding spectral density. The hadronic representation of the spectral density can be obtained by inserting a complete set of corresponding meson states carrying the same quantum numbers as the interpolating current into the correlation function. After straightforward calculations for the spectral densities of the correlation functions from the hadronic side we get
\begin{equation}
  \label{eq:16}
  \rho_\alpha = F_\alpha \delta(s- m_\alpha^2) + F_\alpha^\prime \delta(s-m_\alpha^{\prime2})+...
\end{equation}
where
\begin{equation}
  \label{eq:21}
  F_\alpha =
  \begin{cases}
    \frac{f_{B_c}^{2}(1S) m_{B_c}^4 (1S)}{(m_b + m_c)^2}, m_1 = m_{B_c}(1S)~\text{for } \alpha = 1, \\
    f_{B_c^{*}}^2(1S) m_{B_c^{*}}^2(1S), m_2 = m_{B_c^*}(1S)~\text{for } \alpha = 2, \\
  \end{cases}
\end{equation}
and
\begin{equation}
  \label{eq:22}
  F_\alpha^{\prime} =
  \begin{cases}
    F_\alpha \big(m_{B_c} (1S) \rightarrow m_{B_c} (2S),~f_{B_c}(1S) \rightarrow ~f_{B_c}(2S), m_1^\prime = m_{B_c}(2S)\big) ~\text{for } \alpha = 1, \\ 
    F_\alpha \big(m_{B_c^*} (1S) \rightarrow m_{B_c^*} (2S),~f_{B_c^*}(1S) \rightarrow ~f_{B_c^*}(2S), m_2^\prime = m_{B_c^*}(2S) \big) ~\text{for } \alpha = 2, \\ 
  \end{cases}
\end{equation}
in which $f_{B_c}(1S) (f_{B_c}(2S)), m_{B_c}(1S) (m_{B_c}(2S))$, and $f_{B_c^*}(1S) (f_{B_c^*}(2S)), m_{B_c^*}(1S) (m_{B_c^*}(2S))$ are the leptonic decay constants and mass of the ground (radially excited) pseudoscalar and vector $B_c$ meson states, respectively. In this expression, dots represent the contributions of higher states and continuum. To obtain these spectral densities we used the standard definitions
\begin{equation}
  \label{eq:10}
  \begin{split}
     \langle 0 | J | B_c \rangle &= \frac{f_{B_c} m_{B_c}^2}{m_b + m_c}, \\ 
     \langle 0 | J_\mu| B_c^* \rangle &= f_{B_c^*} m_{B_c^*}\epsilon_\mu
  \end{split}
\end{equation}
where $\epsilon_\mu$ is the $4-$polarization vector meson.
The $\Pi_\alpha(q^2)$ invariant functions can be calculated from local OPE at $q^2 \ll (m_b + m_c)^2$ domain. The OPE part of these correlation functions within QCD sum rules are calculated in~\cite{Colangelo:1992cx,Aliev:1992vp,Wang:2012kw,Bagan:1994dy}. We use the explicit expressions of the correlation functions for vector and pseudoscalar currents presented in \cite{Wang:2012kw} and \cite{Chabab:1993nz}, respectively, in our calculations. Matching the results of the correlation functions from OPE and hadronic parts, we obtain the sum rules for the mass and residues of the pseudoscalar and vector $B_c$-mesons including the contributions of the ground and excited states

\begin{equation}
  \label{eq:5}
  \begin{split}
   \frac{F_\alpha}{m_{\alpha}^2 - q^2} + \frac{F_\alpha^\prime}{m_{\alpha}^{\prime 2} - q^2} + \int_{s_0}^{\infty} \frac{ds}{s-q^2} \rho_\alpha(s) = \int_{(m_b + m_c)^2}^\infty \frac{ds}{s-q^2} \rho_\alpha(s).   
  \end{split}
\end{equation}
In this expression, the third term describes the contributions of higher states, and $s_0$ is the continuum threshold. The hadronic spectral density, according to quark-hadron duality ansatz is equal to the spectral density coming from the OPE part starting from some threshold $s_0$.

Performing Borel transformation over $-q^2$, using quark-hadron duality ansatz and matching two representations of the correlation functions, we get the desired sum rules:
\begin{equation}
  \label{eq:11}
  \begin{split}
F_\alpha e^{-m_\alpha^2/M^2} + F_\alpha^\prime e^{-m_\alpha^{\prime 2}/M^2} = \Pi_\alpha(s_0, M^2)
  \end{split}
\end{equation}
where $\Pi_\alpha(s_0, M^2) = \int_{(m_b + m_c)^2}^{s_0} \rho_\alpha(s) e^{-s/M^2} ds$.
From Eq.~\eqref{eq:11}, we see that these sum rules contain the contributions of the ground \bcones, \bconesp and first radial excited \bctwos and \bctwosp states. In other words, we have one equation and four unknowns (two masses and two residues) for each equation. To determine the masses and leptonic decay constants of the excited $B_c$ meson states, we will use two different approaches within the sum rules. In the first method (Method A), the linear combinations of the corresponding mass sum rules and its derivatives are constructed. On the other hand, the second method (Method B) is based on the corporation of the sum rules with the least squares method where the main idea is minimizing the square of the difference of the hadronic and QCD sides of the correlation function
\begin{equation}
  \label{eq:19}
  \sum_{i=1}^{N} \frac{|F_\alpha e^{-m_\alpha^2/M_i^2} + F_\alpha^\prime e^{-m_\alpha^{\prime^2}/M_i^2} - \Pi_\alpha(M_i^2,s_0)|^2}{N}.
\end{equation}

Choosing a set of $\{ M_i^2 \}$ in the optimal range, we can apply two parameters fitting in order to minimize this function. Both methods have common properties, namely the ground state mass and leptonic decay constants are used as the input parameters. Therefore, at the first stage, we work out the way to reproduce the parameters of the ground state only. It is achieved by choosing the appropriate values of $s_0$, in which the hadronic part contains only ground state pole. 
\section{Numerical Analysis}
\label{sec:3}
Having clarified the details of the methods, let us determine the mass and leptonic decay constants of the ground states \bcones and \bconesp. Even though the mass and residues of the ground states $B_c(1S)$ and $B_c^*(1S)$  (Method A) are estimated in~\cite{Colangelo:1992cx,Aliev:1992vp,Wang:2012kw,Chang:2019eob,Bagan:1994dy} within QCD sum rules, we recalculate these parameters for completeness within Method A. The similar calculations also were performed within method B where the main idea is to check how successful the method B in the predictions of the spectroscopic parameters of $B_c$ meson is. As mentioned earlier, by choosing the appropriate values of $s_0$, the hadronic part can be saturated only by the ground state pole. Hence, the second term in the left side of Eq.\eqref{eq:11} can be omitted. Then applying derivative $\frac{d}{d(-1/M^2)}$ to the both sides of Eq.\eqref{eq:11} we get
\begin{equation}
  \label{eq:17}
F_\alpha m_\alpha^2 e^{-m_\alpha^2/M^2} = \Pi_\alpha^\prime (s_0,M^2)
\end{equation}
where $\Pi_\alpha^\prime = \frac{d \Pi_\alpha(s_0, M^2)}{d(-1/M^2)}$.
Dividing these equations to the left side of Eq.\eqref{eq:11} (without excited state contributions) we obtain
\begin{equation}
  \label{eq:18}
  m_{\alpha}^2 = \frac{\Pi_\alpha^\prime (s_0,M^2)}{\Pi_\alpha(s_0,M^2)}.
\end{equation}

The sum rules contain the auxiliary parameter, namely Borel mass square $M^2$ and continuum threshold, $s_0$. The working regions of $M^2$ for the ground state is determined from the standard criteria; namely, both power corrections and continuum contributions in the sum rules have to be suppressed in this region. The threshold for $s_0$ is chosen in a way that the differentiated sum rules reproduce the measured mass of the ground state meson mass about $10\%$ accuracy.

Numerical analysis shows that these requirements are fulfilled if $M^2$ and $s_0$ vary in the regions presented in Table~\ref{tab:1}. Here, we also depict the working regions of $M^2$ and $s_0$ for the excited states satisfying the conditions mentioned above. Using the values of $M^2$ and $s_0$ from their working regions, we can extract the spectroscopic parameters of the ground state \bcones and \bconesp mesons. The obtained results that are acquired by performing the numerical calculations of Eq.\eqref{eq:11} for the mass and residues of the ground states are presented in Table~\ref{tab:2}.

\begin{table}[h]
  \setlength{\tabcolsep}{12pt}
  \centering
\begin{tabular}{lll}
\hline\noalign{\smallskip}
 & $M^2~(\rm{GeV^2})$ & $s_0~(\rm{GeV^2})$  \\
\noalign{\smallskip}\hline\noalign{\smallskip}
$B_c^+(1S)$ & $5 \le M^2 \leq 15$ & $42 \pm 1$ \\
$B_c^{+*}(1S)$ & $6 \le M^2 \leq 16$ & $43 \pm 1$ \\
$B_c^{+}(2S)$ & $10 \le M^2 \leq 20$ & $53 \pm 1$ \\
$B_c^{+*}(2S)$ & $10 \le M^2 \leq 20$ & $54 \pm 1$ \\
\noalign{\smallskip}\hline
\end{tabular}
  \caption{The working regions of $M^2$ and $s_0$ for the ground $1S$ and excited $2S$ states of $B_c$ mesons.}
  \label{tab:1}
\end{table}

\begin{table}[h]
  \setlength{\tabcolsep}{12pt}
  \centering
\begin{tabular}{lll}
\hline\noalign{\smallskip}
 & Method A & Method B \\
\noalign{\smallskip}\hline\noalign{\smallskip}
$m_{B_c(1S)}$ & $6.28 \pm 0.03$ & $6.28 \pm 0.04$ \\
$\fbcones$ & $0.27 \pm  0.02$ & $0.27 \pm 0.03$ \\
$\mbcones$ & $6.32 \pm 0.03$ & $6.32 \pm 0.04$ \\
$\fbconesp$ & $0.30 \pm  0.02$ & $0.30 \pm 0.03$ \\
\noalign{\smallskip}\hline
\end{tabular}
  \caption{The mass and residues of the ground state \bcones and \bconesp mesons. (In \rm{GeV} units.)}
  \label{tab:2}
\end{table}

\begin{table}[h]
  \setlength{\tabcolsep}{12pt}
  \centering
\begin{tabular}{lll}
\hline\noalign{\smallskip}
 & Method A & Method B  \\
\noalign{\smallskip}\hline\noalign{\smallskip}
$m_{B_c(2S)}$ & $6.88 \pm  0.03$ & $6.88 \pm 0.03$ \\
$f_{B_c(2S)}$ & $0.43 \pm  0.02$ & $0.43 \pm 0.02$ \\
$m_{B_c^*(2S)}$ & $6.94 \pm  0.03$ & $6.95 \pm 0.03$ \\
$f_{B_c^*(2S)}$ & $0.46 \pm  0.01$ & $0.46 \pm 0.02$ \\
\noalign{\smallskip}\hline
\end{tabular}
  \caption{The mass and residues of the excited $B_c^+(2S)$ and $B_c^{*+}(2S)$ meson. (In \rm{GeV} units.)}
  \label{tab:3}
\end{table}

To check the consistency of the methods, we also calculated the mass and residues of the ground states of $B_c$-mesons with the help of Method B. Choosing a set of $\{M_i^2 \}$ in Eq.\eqref{eq:19} in the optimal range, we can apply two-parameters fitting in order to minimize this function. We chose 100 points in the optimal interval of $M^2$. Our results for the mass and residues are also presented in Table~\ref{tab:2}. We observed that both methods lead practically to the same results. 

 Having the mass and residues of the ground state $B_c$-mesons as input parameters, we applied these methods in order to find the mass and residues of the excited states. In this case, we still have to deal with two unknowns. For our first method, we take the derivative with respect to $-1/M^2$ of the Eq.\eqref{eq:11} again and solve for the mass of the excited state. 
 However, note that the range of $s_0$ parameter is different from the ones chosen for the ground state (see Table~\ref{tab:1}). In the method B, the parameters of the ground states are also taken as input ones similar to the method A,  and we try to find the parameter regions of $M^2$ and $s_0$  that minimize the sum of the squares of the difference in Eq.\eqref{eq:19}.

\begin{figure}[bt]
  \centering
  \includegraphics[width=\textwidth]{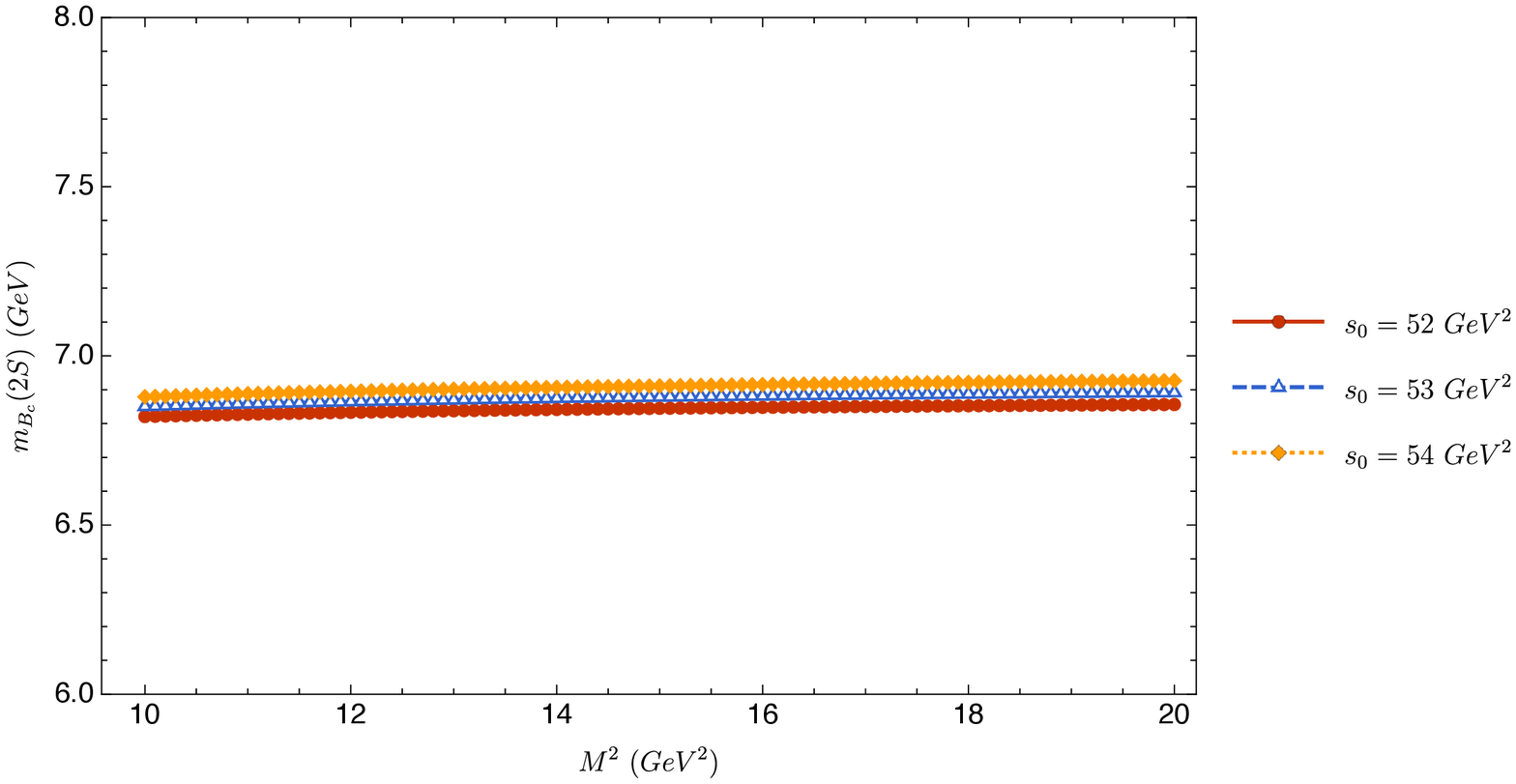}
  \caption{The dependence of the mass $m_{B_c}(2S)$ on Borel mass parameter $M^2$ at three fixed values of $s_0$. }
  \label{fig:fig1}
\end{figure}

\begin{figure}[bt]
  \centering
  \includegraphics[width=\textwidth]{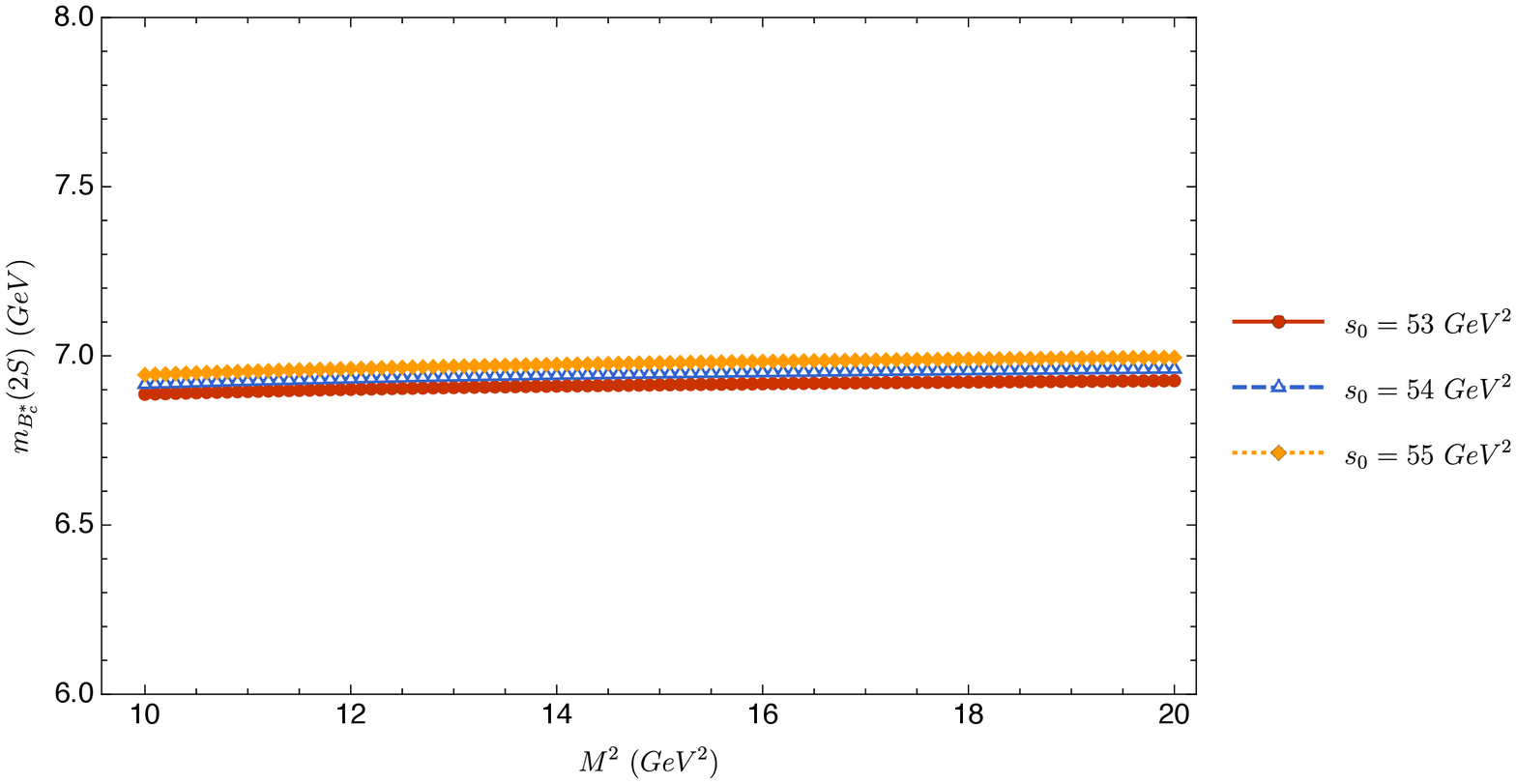}
  \caption{The same as in Fig.~\ref{fig:fig1} but for $m_{B_c^*}(2S)$.}
  \label{fig:fig2}
\end{figure}

In \cref{fig:fig1,fig:fig2}, we present the results of the numerical calculations of the  dependence of the mass of \bctwos and \bctwosp as a function of $M^2$ at three fixed values of $s_0$ within method A. From these figures, it follows that the mass of the excited states exhibits good stability to the variation of $M^2$ from its working region. Our final predictions for the mass of \bctwosp and \bctwos states within both methods are presented in Table~\ref{tab:3}.

\begin{figure}[bt]
  \centering
  \includegraphics[width=\textwidth]{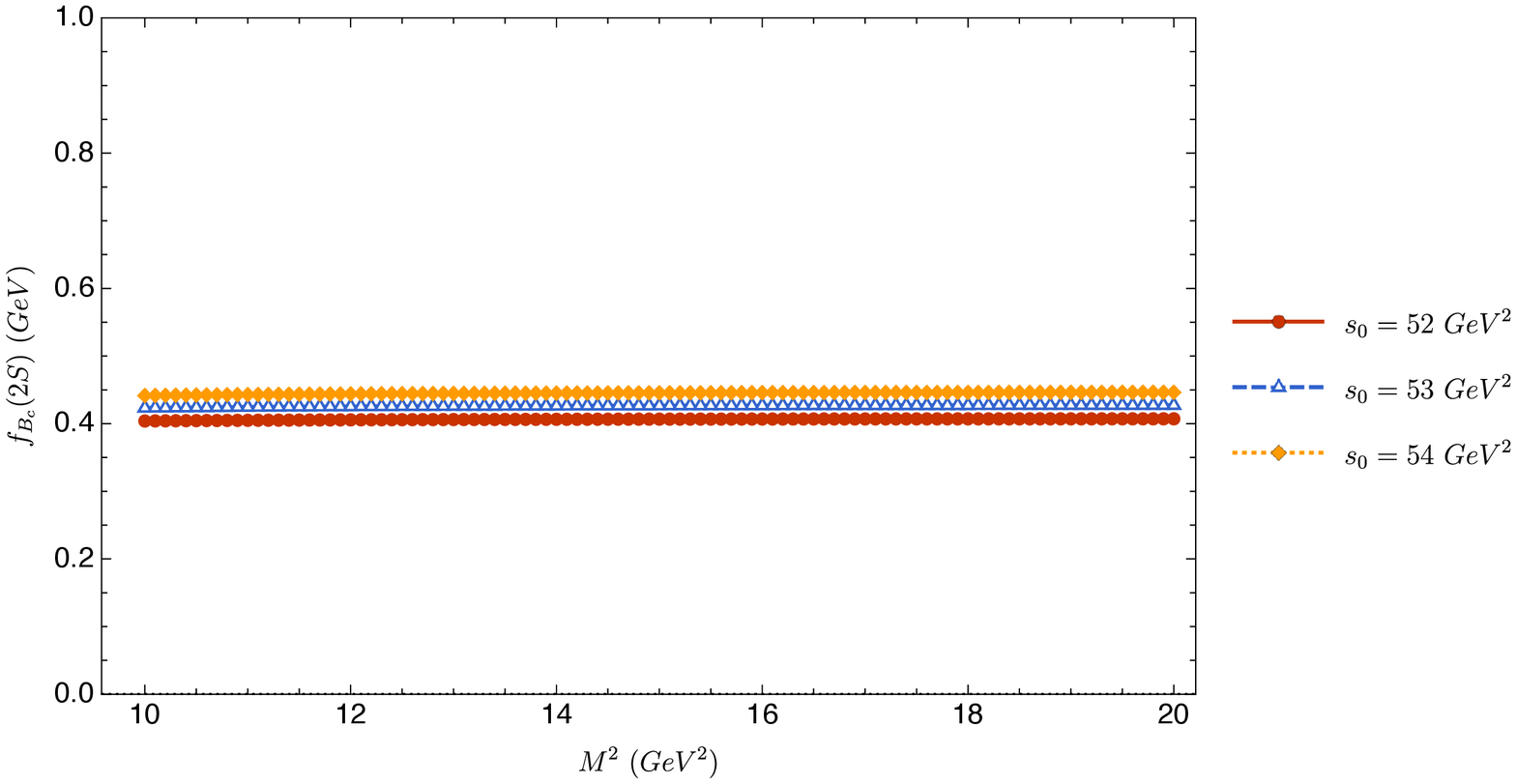}
  \caption{The dependence of the leptonic decay constants of $B_c(2S)$ on Borel mass parameter $M^2$ at three fixed values of $s_0$.}
  \label{fig:fig3}
\end{figure}

\begin{figure}[bt]
  \centering
  \includegraphics[width=\textwidth]{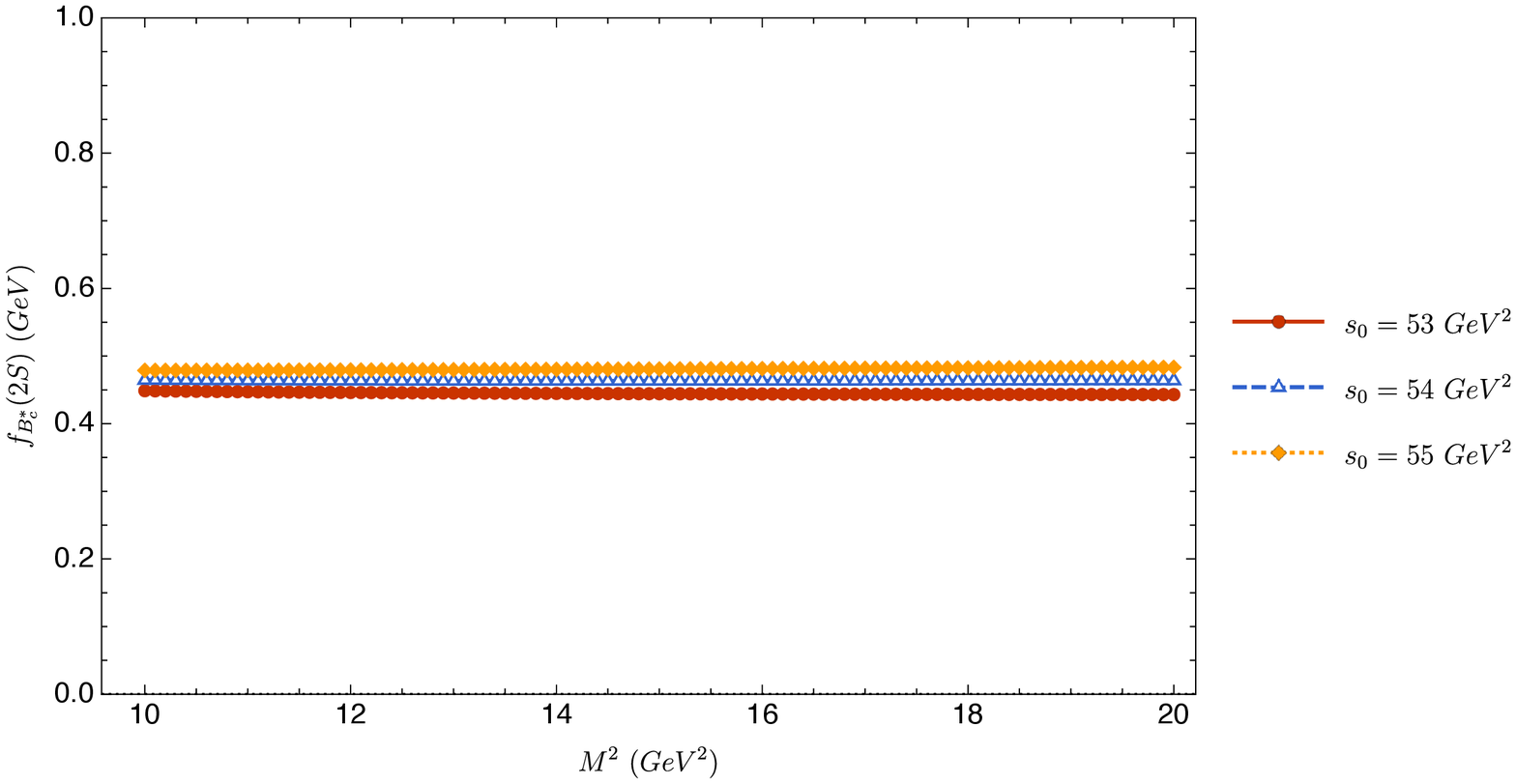}
  \caption{The same as in Fig.~\ref{fig:fig3} but for $B_c^*(2S)$}.
  \label{fig:fig4}
\end{figure}

In \cref{fig:fig3,fig:fig4}, the dependence of the leptonic decay constants of the \bctwos and \bctwosp states on $M^2$ at three fixed values of $s_0$ are presented in the framework of the method A. Moreover, we also calculated the leptonic constants of the excited $2S$ states within method B. Our final results on this quantity are given in Table~\ref{tab:3}.

A glance of our predictions on the mass of the excited $B_c$ shows that our result is nicely in agreement with the experimental results. Besides, our findings on the mass difference $\Delta m \cong 30~\rm{MeV}$ is also in very good agreement with the experimental observation.

From the results stated in Table~\ref{tab:2} and \ref{tab:3}, we observe that $f_{B(2S)} > f_{B(1S)}$ and $f_{B^*(2S)} > f_{B^*(1S)}$. These results indicated that the currents might have a large overlap with the excited states. This point needs further detailed investigation. It should be noted that similar results are observed for $Z_c(3900)$ and its radial excitations~\cite{Wang:2014vha,Wang:2019hnw}. 

~At~ the end of this section, we compare our results on mass and residues of the excited $B_c$ meson states obtained in the framework of the Dyson-Schwinger and Bethe-Salpeter equation approaches of continuum QCD~\cite{Chang:2019eob}.
\begin{equation}
  \label{eq:9}
  \begin{split}
    \mbctwos &= 6.813~\rm{GeV} \\
    \mbctwosp &= 6.84~\rm{GeV} \\
    \fbctwos &= -0.165~\rm{GeV} \\
    \fbctwosp &= -0.161~\rm{GeV} 
  \end{split}
\end{equation}

Comparing these results with our predictions, we observe that they are in good agreement for the mass of the excited states. However, the results for the leptonic decay constants are considerably different. Future experiments can shed light into this discrepancy.

\section{Conclusion}
The two excited $B_c$ meson states have been observed by LHC very recently. In this study, we calculated the mass and decay constants of these states through the QCD sum rule method. The obtained results on mass and mass difference are in good agreement with the experimental data. Moreover, we predicted the residues of the excited $B_c$ mesons, which are considerably different from the one predicted in~\cite{Chang:2019eob}. Our result on leptonic decay constants can be checked at future experiments.

\section{Acknowledgments:}
One of the authors, T. Barakat, thanks to International Scientific Partnership Program ISPP at the King Saud University for funding his research work through ISPP No:0038.

\bibliography{B_c_mesons}

\newpage

\end{document}